\documentclass[twocolumn,twoside,slac_two]{revtex4}
\usepackage{graphicx}
\usepackage{fancyhdr}
\begin{document}

%Title of paper
\title{Measurement of \boldmath{$B^{+}_{c}$} properties at CDF}

% Repeat the \author .. \affiliation  etc. as needed
%
% \affiliation command applies to all authors since the last
% \affiliation command. The \affiliation command should follow the
% other information

\author{T.S. Nigmanov$^{a}$\footnote{Speaker, on behalf of the CDF 
Collaboration}, K.R. Gibson$^{b}$, M.P. Hartz$^{c}$, 
P.F. Shepard$^{b}$}
\affiliation{$^{a}$University of Michigan, Ann Arbor, MI 48109, USA} 
\affiliation{$^{b}$University of Pittsburgh, Pittsburgh, PA 15260, USA} 
\affiliation{$^{c}$University of Toronto, Toronto M5S, Canada}

\begin{abstract}
The $B^{+}_{c}$ meson is composed of two heavy quarks of distinct flavor. 
Measurements of its lifetime and production properties have been made based 
on semileptonic $B^{+}_{c}\to
J/\psi+l^{+}+X$ decays using data collected with the CDF\,II detector 
corresponding to an integrated luminosity of 1 fb$^{-1}$. The $B^{+}_{c}$
average lifetime $c\tau$ is measured to be 
142.5$^{+15.8}_{-14.8}$(stat)$\pm 5.5(syst)$ $\mu m$. The measurements of the 
ratio of the production cross section times branching ratio of $B^{+}_{c}\to 
J/\psi\mu^{+}\nu$ relative to $B^{+}\to J/\psi K^{+}$ were done for two 
$p_{T}(B)$ thresholds: for $p_T(B)>$ 4 GeV/$c$ as $0.295\pm 
0.040~\mbox{(stat.)}  ^{+0.033}_{-0.026}~\mbox{(syst.)} 
\pm0.036~(p_{T}~\mbox{spectrum})$ and for $p_T(B)>$ 6 GeV/$c$ as 
$0.227\pm 0.033~\mbox{(stat.)}  ^{+0.024}_{-0.017}~\mbox{(syst.)} 
\pm0.014~(p_T~\mbox{spectrum})$.
\end{abstract}

%\maketitle must follow title, authors, abstract
\maketitle

\thispagestyle{fancy}

% body of paper here - Use proper section commands
% References should be done using the \cite, \ref, and \label commands
% Put \label in argument of \section for cross-referencing
%\section{\label{}}

%%%%%%%%%%%%%%%%%%%%%%%%%%%%%%%%%%
\section{\label{sec:Intro}Introduction}
The $B^{+}_{c}$ meson ~\cite{charge_conjugate} is composed of an anti-bottom 
quark $\bar{b}$ and a charm quark $c$. The presence of two relatively
heavy quarks with different flavors is unique to the $B^{+}_{c}$ system and 
affects the decay and production properties. The theoretically predicted 
lifetime ~\cite{kiselev} is about a factor of three times smaller than that 
of other B mesons. The expected $B^{+}_{c}$ production cross section 
~\cite{chang} is about 3 orders of magnitude lower than the production cross 
section of the $B^{+}$ ~\cite{CDF-bplus-Xsect}. The first observation of the 
$B^{+}_{c}$ was made using data taken with the CDF detector at the Fermilab 
Tevatron during run I~\cite{bc-observation}. Precise mass measurements have 
been made by the CDF Collaboration using fully reconstructed 
$B^{+}_{c}\rightarrow J/\psi\pi^{+}$ decays,
where $J/\psi$ decays through $J/\psi\rightarrow\mu^{+}\mu^{-}$ 
~\cite{bc-mass}. 

In this work we report preliminary measurements of the $B^{+}_{c}$ lifetime 
in the semileptonic decay modes $J/\psi \mu^+ X$ and $J/\psi e X$, and the 
production cross section times branching ratio of the decay mode 
$B^{+}_{c}\rightarrow J/\psi\mu^{+} \nu$ relative to the 
$B^{+}\rightarrow J/\psi K^{+}$ decay. The results presented here
are based on a data sample  with an integrated luminosity of 1 fb$^{-1}$ 
at $\sqrt{s}$=1.96 ~TeV collected by the CDF\,II detector.

%%%%%%%%%%%%%%%%%%%%%%%%%%%%%%%%%%
\section{\label{sec:lifeApproach} The \boldmath{$B^{+}_{c}$} lifetime 
measurement concept}

To measure the lifetime of the $B^{+}_{c}$, we construct a per event
lifetime that is defined using variables measured in the transverse
plane.  If all of the decay products of the $B^{+}_{c}$ decay are
identified, the lifetime $ct$ is the lifetime of the
$B^{+}_{c}$~meson in its rest frame measured in units of microns of light
travel time. It is expressed as
\begin{equation}
ct = \frac{m L_{xy}}{p_T}
\end{equation}
where $m$ is the mass of the $B^{+}_{c}$, $p_T$ is the momentum of the 
$B^{+}_{c}$ in the plane transverse to the direction of the proton beam, 
and $L_{xy}$ is the decay length of the $B^{+}_{c}$ projected along the
transverse momentum.  The mass of the $B^{+}_{c}$ used in this measurement
is $m=6.286$ GeV/c$^{2}$~\cite{bc-mass}.  However, we do not measure all of
the particles in the semileptonic $B^{+}_{c}$ final state.  Instead we must 
define a pseudo lifetime
\begin{equation}
{ct^* = \frac{m L_{xy}(J/\psi l^{+})}{p_T(J/\psi l^{+})}}
\end{equation}
where $L_{xy}$ and $p_T$ are evaluated using the $J/\psi+l^{+}$ system.
We can obtain the true $B^{+}_{c}$ lifetime by defining a factor $K$, where
$ct = Kct^*$.  We evaluate the $K$ factor distribution, $H(K)$, for
$B^{+}_c$ events using Monte Carlo simulation.  We are then able to express 
the  distribution of $ct^*$ for $B^{+}_c\rightarrow J/\psi l^{+} X$ as
\begin{equation} 
{F_{B_c}(ct^*,\sigma)=\int dK H(K)\frac{K}{c\tau}\theta(ct^*)exp(-\frac{Kct^*}{c\tau}) \otimes G(\sigma)}
\end{equation}
where $c\tau$ is the average $B^{+}_{c}$ lifetime, $\sigma$ represents
the estimated error on the measurement of $ct^*$ for each event, and 
$G(\sigma)$ is defined as
\begin{equation}  
{G(\sigma) = \frac{1}{\sqrt{2\pi}s\sigma}e^{-\frac{1}{2}(\frac{ct^*}{s\sigma})^2} }
\end{equation}
 
The measurement of the $B^{+}_c$ average lifetime is be carried out by
minimizing $-2Log(L)$, which is evaluated for the candidate $J/\psi+l$
events.  $L$ is the likelihood function for the $ct^*$ and $\sigma$
measured in candidate events and includes $c\tau$ as a free
parameter.

%%%%%%%%%%%%%%%%%%%%%%%%%%%%%%%%%%
\section{\label{sec:crossApproach} The \boldmath{$B^{+}_{c}$} cross section 
measurement concept}

We measure 
\begin{equation} 
{\frac{\sigma(B^{+}_c)BF(B^{+}_c\to J/\psi\mu^{+}\nu)}{\sigma(B^{+})BF(B^{+}\to J/\psi K^{+})} = \frac{N(B^{+}_c)}{N(B^{+})}\times\epsilon_{rel}}
\end{equation}

The basic strategy of the $B^{+}_{c}$ cross section measurement is to 
reconstruct the number of $B^{+}_c\to J/\psi\mu^{+}\nu$ relative to the 
number of $B^{+}\to J/\psi K^{+}$ candidates, $N(B^{+}_c)/N(B^{+})$, 
determine the relative detector and reconstruction efficiency, 
$\epsilon_{rel}$ = $\epsilon(B^{+})/\epsilon(B^{+}_c)$, and use these to 
determine the ratio of the final production cross section times branching 
ratio.

%%%%%%%%%%%%%%%%%%%%%%%%%%%%%%%%%%
\section{\label{sec:eventSelection} The event selection}

\begin{figure}[h]
\centering
\includegraphics[width=60mm]{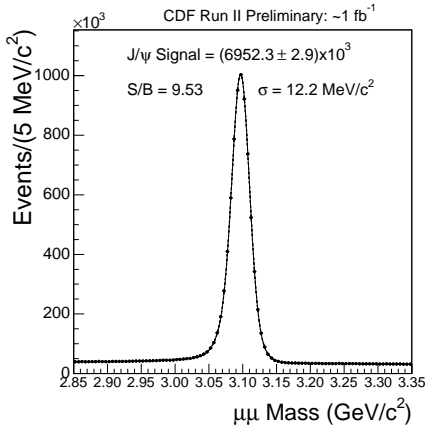}
\caption{The $J/\psi\rightarrow \mu^+\mu^-$ invariant mass distribution.  
Events within a mass range of $\pm$50 MeV/$c^{2}$ around the central 
$J/\psi$ mass value were used for both the lifetime and cross section 
analysis.} 
\label{jpsi_mass}
\end{figure}

The analysis presented here is based on the events recorded with a di-muon
trigger that is dedicated to $J/\psi\to\mu^{+}\mu^{-}$ decays. Both 
analysis use the same $J/\psi\to\mu^{+}\mu^{-}$ sample.  The muon pair is 
reconstructed within a pseudo-rapidity range $|\eta|<$ 1.0. We select about 
6.9$\times$10$^{6}$ $J/\psi$ candidates, 
measured with a mass resolution of approximately 12 MeV/$c^{2}$. The di-muon 
invariant mass distribution is shown in Fig.~\ref{jpsi_mass}.
  
In addition to the di-muons from the $J/\psi$ decay, we require a third track 
that is matched to the same vertex as the $J/\psi$.  This third track can 
be from any of three samples of interest: $B^{+}_{c}\to J/\psi l^{+}$X decays, 
$B^{+}\to J/\psi K^{+}$ decays, or just $J/\psi + track$ decays. The 
last sample represents sources of backgrounds to the 
$B^{+}_{c}$ semileptonic decay.

%%%%%%%%%%%%%%%%%%%%%%%%%%%%%%%%%%
\section{\label{sec:backgrounds} The \boldmath{$B^{+}_{c}$} backgrounds
overview}

\subsection{\label{sec:backgrounds-common} Common for both analysis}

Both the lifetime and cross section analysis have some common sources of
background: misidentified $J/\psi$, misidentified third muons, and 
$b\bar{b}$ backgrounds. The misidentified $J/\psi$ background occurs when 
one of the muons is actually a mis-reconstructed hadron or muon from other 
sources that produce a mass consistent with that of the $J/\psi$. The 
misidentified third muon background can arise from the following sources.
The $J/\psi$ in the $J/\psi+track$ system is highly populated by 
non-$B^{+}_{c}$ sources. The third track associated with the $J/\psi$ could 
be a $\pi^{+}$ or $K^{+}$ that can either decay-in-flight 
or punch-through the calorimeter and the steel absorber and produce the muon 
signature.  The $b\bar{b}$ events represent cases when a $J/\psi$ is produced 
from one $b$ jet and the third muon originates from the other $b$ in same 
event. 
  
\begin{figure}[h]
\centering
\includegraphics[width=40mm]{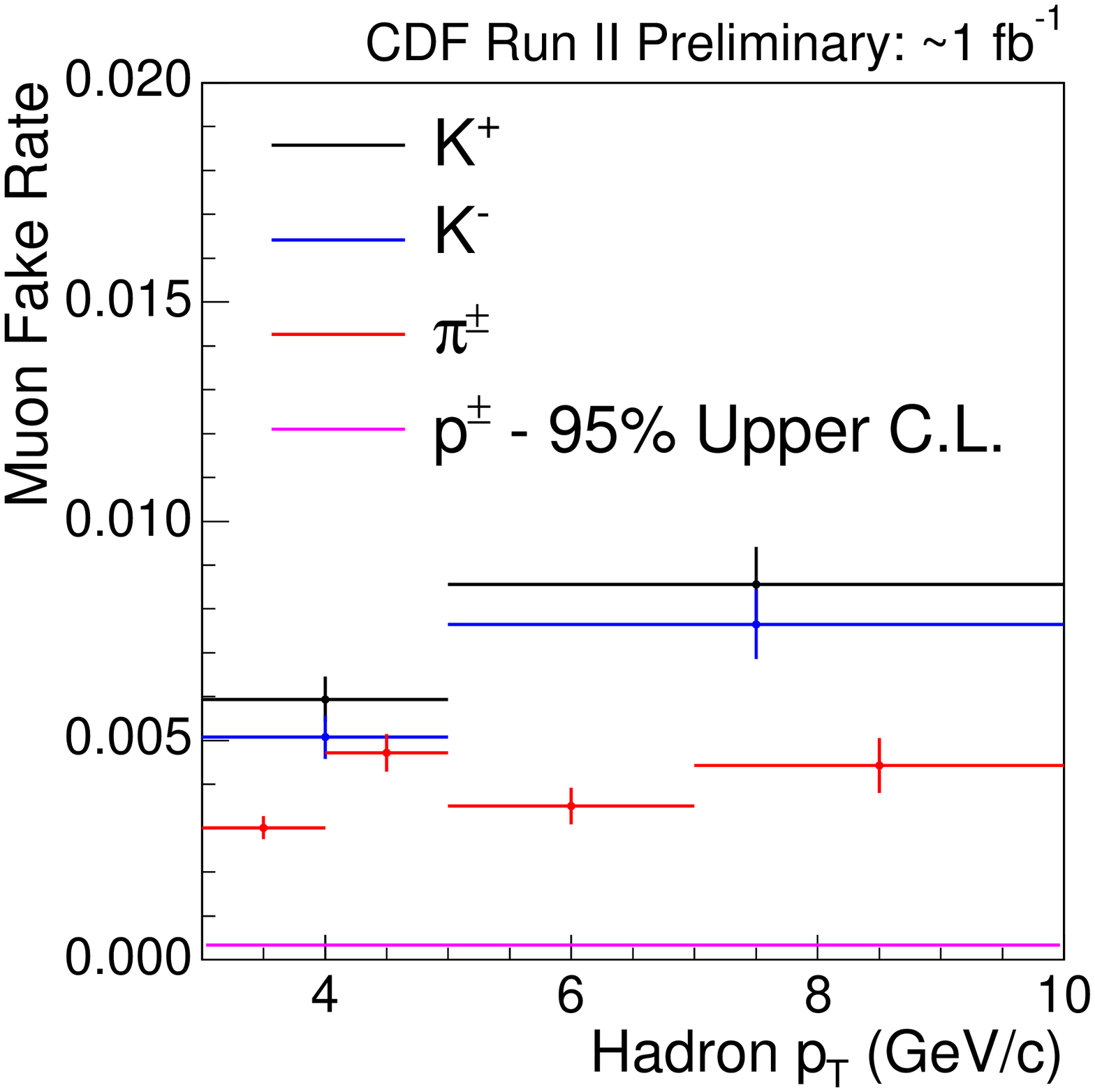}
\includegraphics[width=40mm]{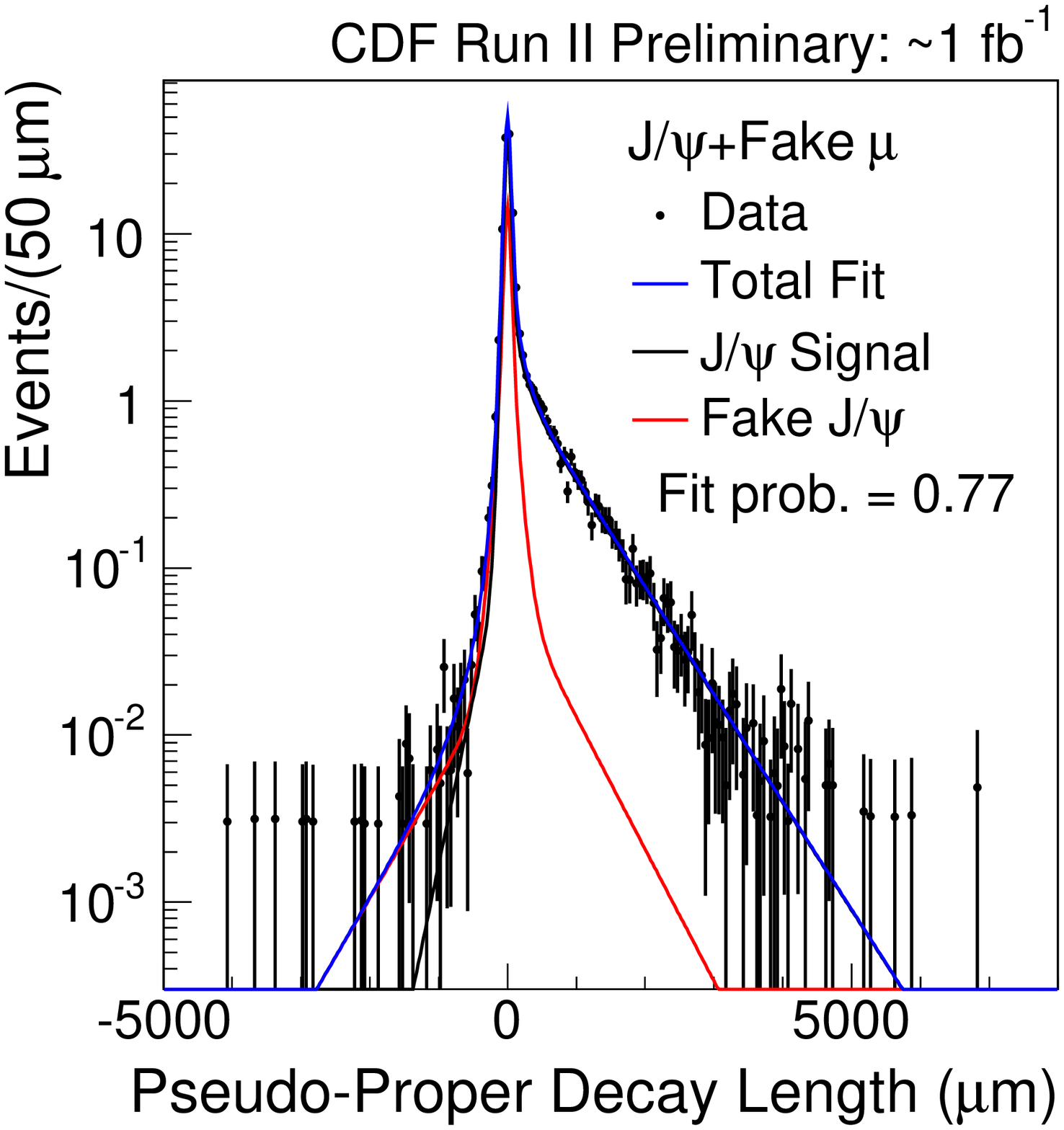} \\
\caption{The misidentified muon rates from $\pi^{\pm}$, $K^{\pm}$, and 
p($\bar{p}$) as a function of hadron $p_T$ (left), and the pseudo-proper 
decay length ct$^{*}$ for misidentified third muons (right).} 
\label{fakeProb_fakeMuLife}
\end{figure}

Figure~\ref{fakeProb_fakeMuLife} shows the misidentified muon rates for 
$\pi^{\pm}$, $K^{\pm}$, and p($\bar{p}$) and the pseudo-proper 
decay length ct$^{*}$ for misidentified third muons. 

%%%%%%%%%%%%%%%%%%%%%%%%%%%%%%%%%%
\subsection{\label{sec:backgrounds-specific} Specific for each analysis}

There is an additional background for the cross section measurement.
The selected $B^{+}_{c}\to J/\psi\mu^{+}X$ sample contains contributions 
from other $B^{+}_{c}$ decays with a tri-muon in the final state. For example, 
a $B^{+}_{c}$ can decay into $\psi(2S)\mu^+\nu$ followed by 
$\psi(2S)\rightarrow J/\psi X$.

\begin{figure}[ht]
\centering
\includegraphics[width=40mm]{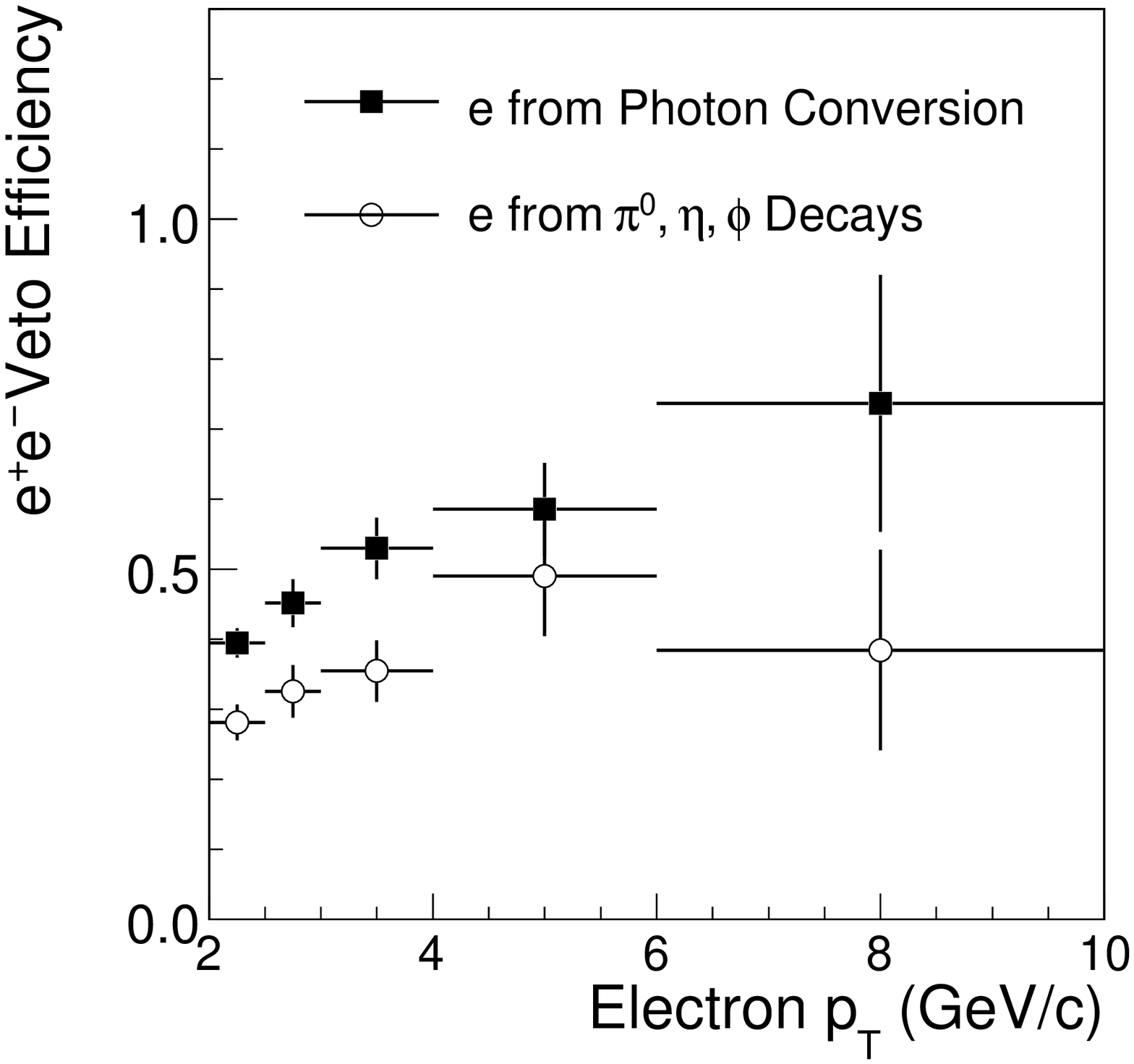}
\includegraphics[width=40mm]{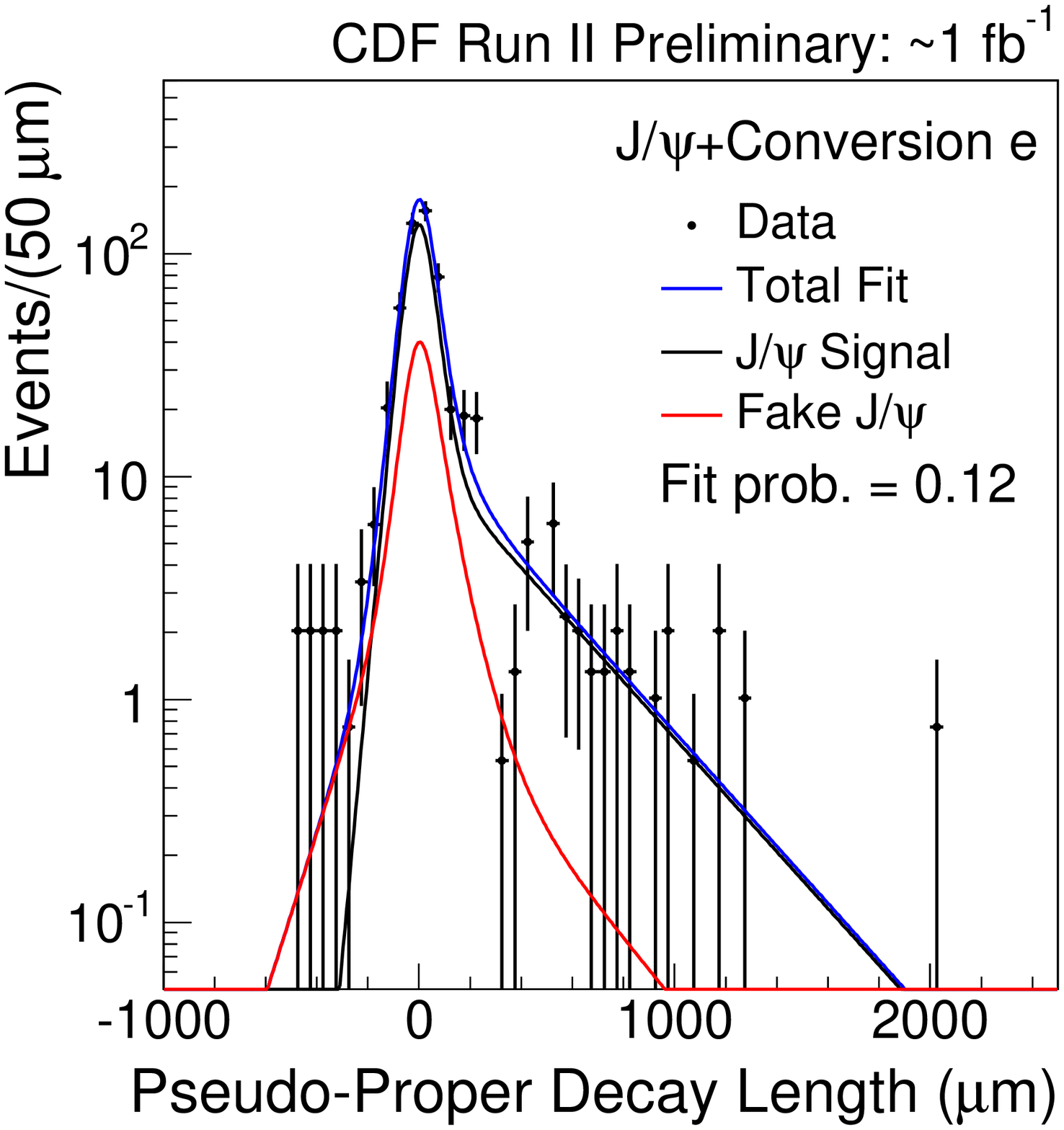} \\
\caption{The $e^{+}e^{-}$ veto efficiencies as a function of electron $p_T$
(left), and the pseudo-proper decay length ct$^{*}$ distribution for the  
$J/\psi$+Conversion e background sample (right).} 
\label{convProb_convLife}
\end{figure}

The following backgrounds are specific to the lifetime analysis: misidentified 
e$^{\pm}$, residual conversions, and prompt $J/\psi$. The misidentified
e$^{\pm}$ can arise from cases when a $\pi^{\pm}$, K$^{\pm}$, or $\bar{p}$ 
from the $J/\psi+track$ system satisfies the e$^{\pm}$ likelihood function 
based on the calorimeter responses. The residual conversions are e$^{\pm}$ 
from $\gamma$-conversion or $\pi^{o}$ Dalitz decays. The prompt $J/\psi$ 
are additional $J/\psi l^{\pm}$ candidates where the $J/\psi$ originates from
prompt non-$B^{+}_{c}$ sources. 
Figure~\ref{convProb_convLife} illustrates the $e^{+}e^{-}$ veto efficiencies
and the pseudo-proper decay length ct$^{*}$ distribution for the  
$J/\psi$+Conversion e background sample. 

%%%%%%%%%%%%%%%%%%%%%%%%%%%%%%%%%%
\section{\label{sec:lifetime} The \boldmath{$B^{+}_{c}$} lifetime 
results}

%%%%%%%%%%%%%%%%%%%%%%%%%%%%%%%%%%
\subsection{\label{sec:lifetime_sys} Lifetime systematic uncertainties}

The systematic uncertainties in the $B^{+}_c$ lifetime measurement
originate from uncertainties in our models for background and signal
events.  Some of the largest systematic uncertainties are summarized below:

\begin{itemize}
\item{Resolution function - choice of model for detector resolution: $\pm$3.8 
$\mu m$}
\item{Pythia model for $b\bar{b}$ background - relative contribution
of QCD processes: $\pm$2.4 $\mu m$}
\item{Vertex detector alignment - uncertainties in the positions of  silicon 
detectors: $\pm$2.0 $\mu m$}
\item{$e^{+}e^{-}$ veto efficiency - uncertainties related to modeling 
$e^{+}e^{-}$ veto efficiencies: $\pm$1.5 $\mu m$}
\item{$B_{c}$ spectrum - variations of the K factor distribution due to
variations in the $B_{c}$ production spectrum: $\pm$1.3 $\mu m$}
\end{itemize}

We add the individual uncertainties in  quadrature to obtain a total 
uncertainty of $\pm$5.5 $\mu m$.

%%%%%%%%%%%%%%%%%%%%%%%%%%%%%%%%%%
\subsection{\label{sec:lifetime_results} Lifetime results}

\begin{figure}[ht]
\centering
\includegraphics[width=40mm]{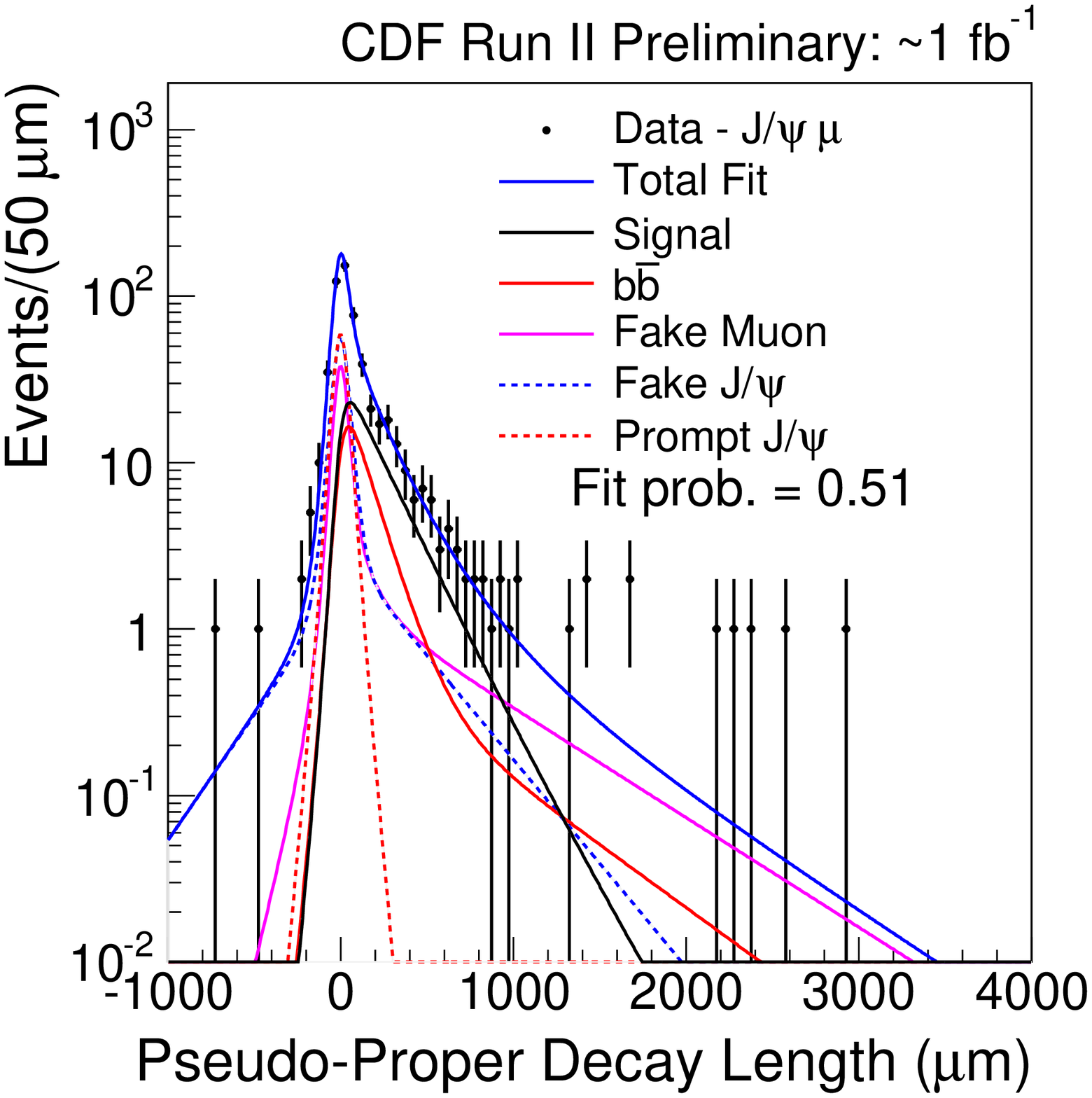}
\includegraphics[width=40mm]{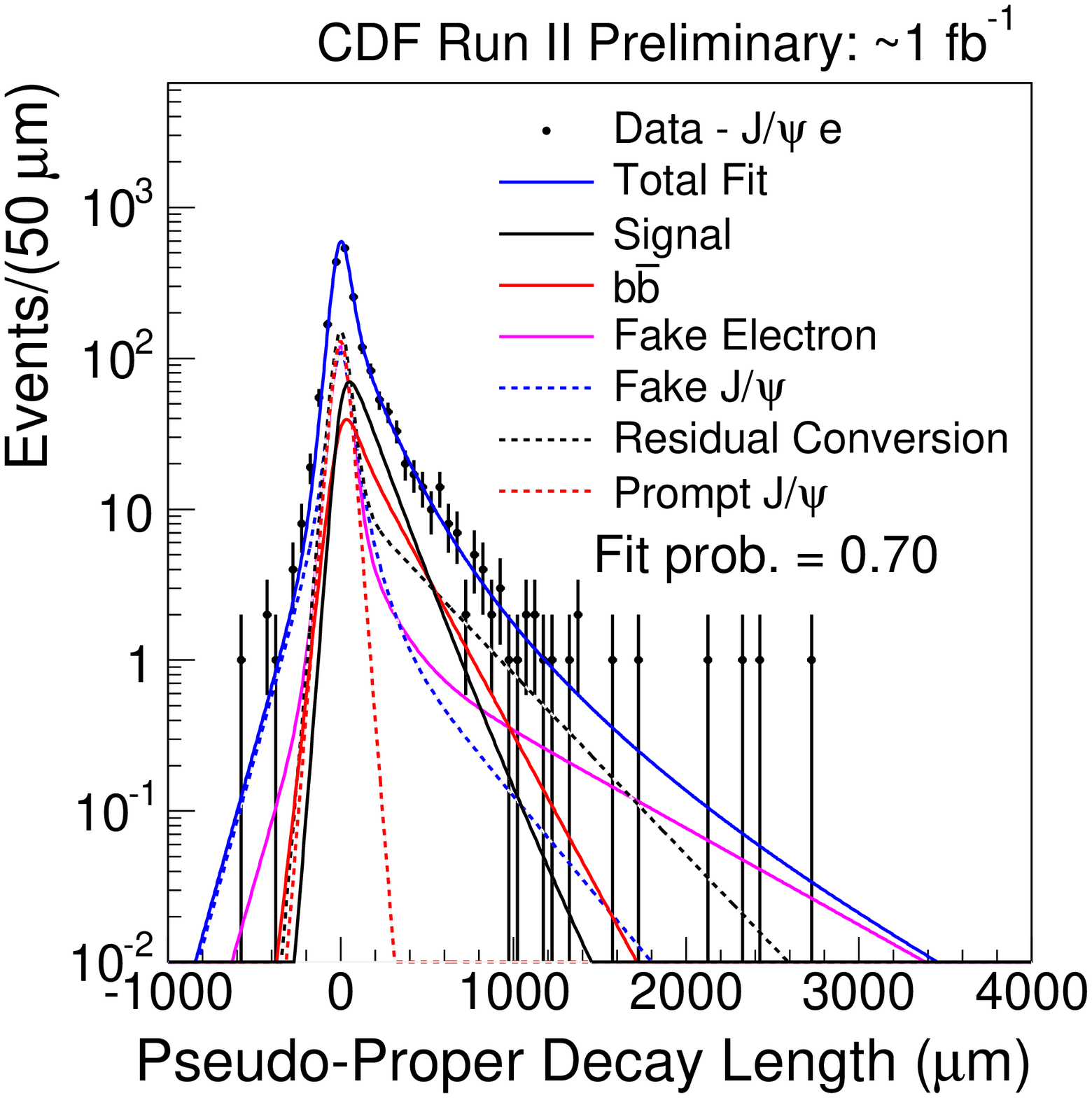} \\
\caption{The pseudo-proper decay length ct$^{*}$ distributions for 
$B^{+}_{c}\to J/\psi\mu^{+}X$ decays (left) and for  $B^{+}_{c}\to J/\psi 
e^{+}X$ decays (right) with their the background models superimposed.} 
\label{bc_lifetime_plots}
\end{figure}

We fit the $ct^*$ distributions for signal candidates in the
$J/\psi\mu^{+}$ and $J/\psi e^{+}$ channels separately using likelihood
functions based on our models for signal and background events.  
The fitted data is shown in Fig.~\ref{bc_lifetime_plots} for 
$B^{+}_{c}\to J/\psi\mu^{+}X$ and for  
$B^{+}_{c}\to J/\psi e^{+}X$ decays.

\begin{figure}[h]
\centering
\includegraphics[width=60mm]{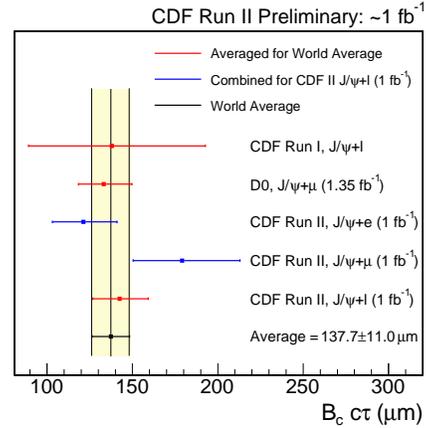}
\caption{World average of $B^{+}_c$ lifetime, which includes the CDF 
Run\,I $B_c$ lifetime, the most recent D0 Run\,II result, and
the result presented in this paper.  The lifetimes are weighted by the
total variance of the individual measurements in the average.} 
\label{lifetime_world_avg}
\end{figure}

The $B^{+}_{c}$ lifetime is found to be 179.1$^{+32.6}_{-27.2}$(stat) 
$\mu m$ for the $J/\psi\mu^{+}$ final state and 121.7$^{+18.0}_{-16.3}$
(stat) $\mu m$ for the $J/\psi e^{+}$ decay mode, respectively.
We performed the simultaneous fit of both samples and found an average
$B^{+}_{c}$ lifetime of 142.5$^{+15.8}_{-14.8}(stat)\pm5.5(syst)$ $\mu m$.
Figure~\ref{lifetime_world_avg} shows our $B^{+}_{c}$ average lifetime
comparison with other measurements.

%%%%%%%%%%%%%%%%%%%%%%%%%%%%%%%%%%
\section{\label{sec:bc_Xsec_all} The \boldmath{$B^{+}_{c}$} relative 
cross section}

In order to measure the $B^{+}_{c}$ relative cross section we need to 
find the numbers of $B_c^+$ and $B^+$, $N(B^{+}_c$) and $N(B^{+}$), and  
determine the relative efficiency, $\epsilon_{rel}$ = 
$\epsilon(B^{+})/\epsilon(B^{+}_c)$.
We select 229 (214) $B^{+}_{c}$ candidates with the requirement 
$p_T(B^{+}_{c})>$ 4 (6) GeV/$c$, respectively. The number of $B^{+}_{c}$ 
signal events after backgrounds subtraction is presented in the following 
subsection.  The number of $B^{+}\to J/\psi K^{+}$ signal events is found 
to be 2333 $\pm$ 55 (2299 $\pm$ 53) for $p_T(B^{+})>$ 4 (6) GeV/$c$, 
respectively. The combinatoric and
$B^{+}\to J/\psi\pi^+$ contributions are subtracted.

%%%%%%%%%%%%%%%%%%%%%%%%%%%%%%%%%%
\subsection{\label{sec:bc_Xsec_signal} The \boldmath{$B^{+}_{c}$} 
backgrounds and excess}

The backgrounds and the resulting number of signal events for the 
$B^{+}_{c}\to J/\psi\mu^{+}\nu$ decays are summarized in 
Table~\ref{tab_bc_excess}.
\begin{table}[htbp]
\begin{center}
%\label{tab_bc_excess}
\caption{Observed $N(B^{+}_{c}\to J/\psi\mu^{+}\nu$) for the 
$p_T(J/\psi\mu)>$ 4 GeV/c (6 GeV/c) threshold.}
\label{tab_bc_excess}
\begin{tabular}{|l|c|c|}
\hline  

                         & $p_{T}(B)>$ 4 GeV/$c$ & $p_{T}(B)>$ 6 GeV/$c$ \\
\hline
$N(B^{+}_{c}$) observed  & 229$\pm$15.1(stat) & 214$\pm$14.6(stat) \\
\hline
Misidentified $J/\psi$   & 21.5$\pm$3.3(stat) & 20.5$\pm$3.2(stat) \\
\hline
Misid. third muon        & 55.8$\pm$2.0(stat) & 53.6$\pm$1.9(stat) \\
\hline
Doubly misid.            &-8.8$\pm$0.4(stat)  & -7.5$\pm$0.3(stat) \\
\hline
$b\bar{b}$ background    &37.7$\pm$7.3(st+sys)& 35.4$\pm$7.0(st+sys)\\
\hline
Other $B^{+}_{c}$ modes  & 5.2$\pm$0.5(stat)  &  4.8$\pm$0.4(stat)  \\
\hline
Total background         & 111.4$\pm$8.3(stat)& 106.9$\pm$8.0(stat)\\
\hline
$B^{+}_{c}$ signal       & 117.6$\pm$17.2(stat)& 107.1$\pm$16.7(stat) \\
\hline
\end{tabular}
\end{center}
\end{table}
The background identified as ``Doubly misidentified'' in 
Table~\ref{tab_bc_excess} 
represents the subsample of misidentified $J/\psi$ and misidentified third 
muons that needs to be subtracted only once to avoid double counting.

\begin{figure}[ht]
\centering
\includegraphics[width=40mm]{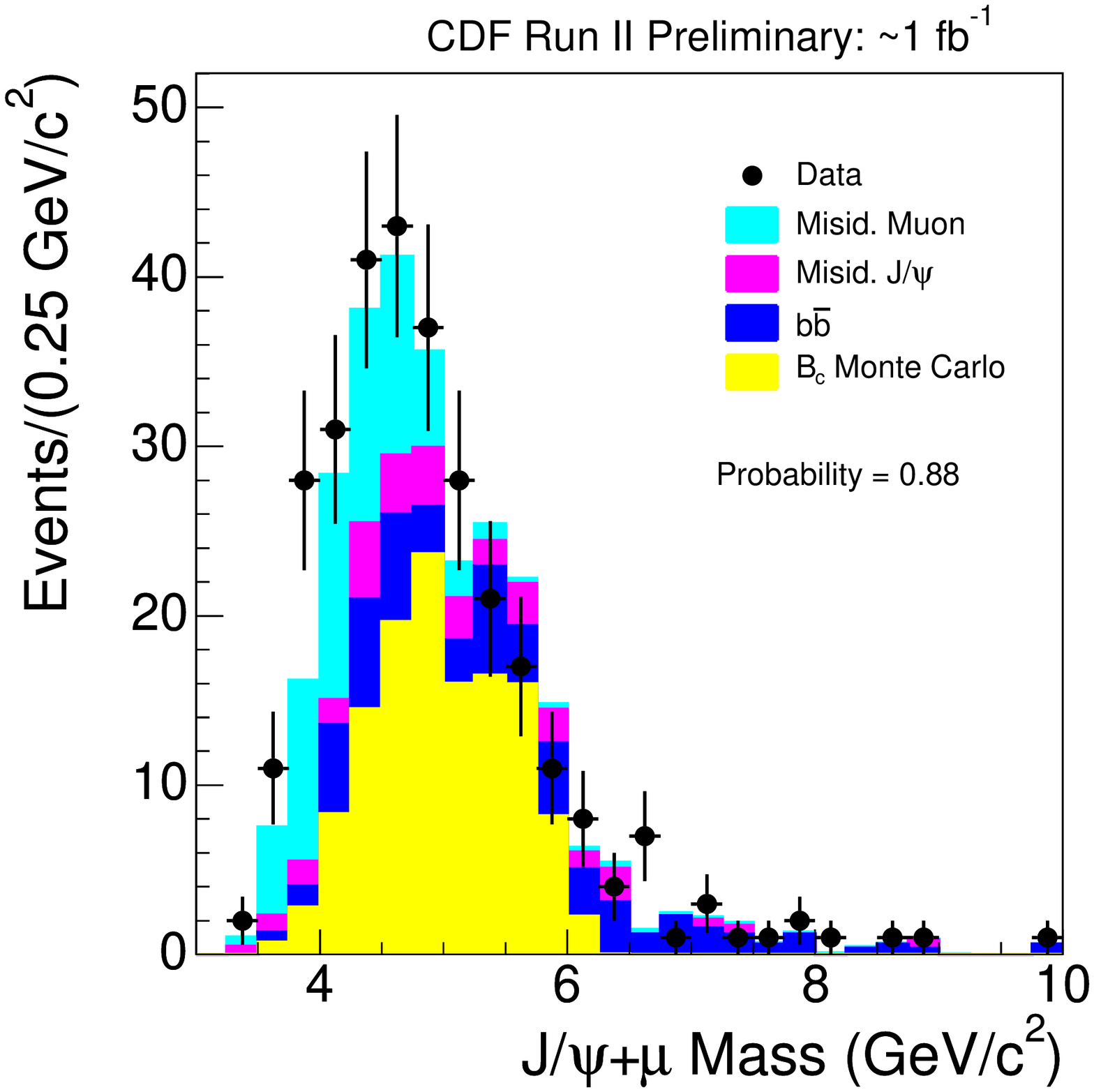}
\includegraphics[width=40mm]{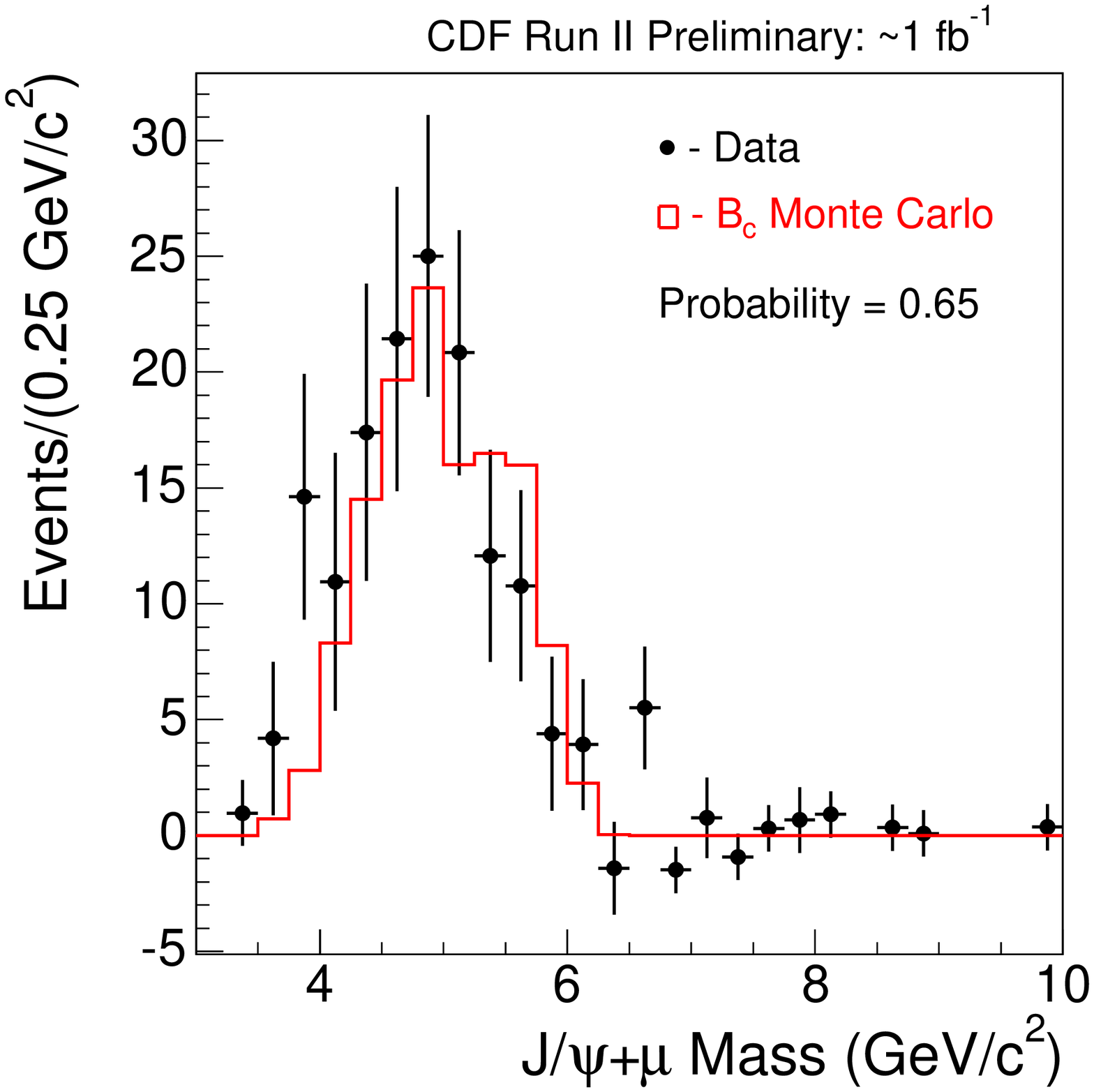} \\
\caption{The invariant mass distribution of the $B^{+}_{c}\to J/\psi\mu^{+}X$ 
data events with the experimental backgrounds and a Monte Carlo simulation of 
the signal sample superimposed (left), and with the background subtracted 
applied (right).} 
\label{bc_mass_plots}
\end{figure}

Figure~\ref{bc_mass_plots} shows the invariant mass distribution of 
the $B^{+}_{c}\to J/\psi\mu^{+}X$ data events with the experimental 
backgrounds and a Monte Carlo simulation of the signal sample superimposed 
(left), and with the background subtracted (right).

%%%%%%%%%%%%%%%%%%%%%%%%%%%%%%%%%%
\subsection{\label{sec:bc_Xsec_erel} The relative efficiency 
\boldmath{$\epsilon_{rel}$}}

In order to determine the relative efficiency, we simulate $B^{+}\to 
J/\psi K^{+}$, $B^{+}_{c}\to J/\psi\mu^{+}\nu$, and 
$B^{*+}_{c}\to B^{+}_c\gamma$ decays.  As a description of the $\eta-p_T$ 
spectrum for $B^{+}_c$ we use the most recent theoretical work by Chang {\it 
et al.}~\cite{chang}.  For the $B^{+}$ we used the spectrum from 
Ref.~\cite{FONLL}, which is found to be in good agreement with CDF 
measurements.  All Monte Carlo simulation events are passed through the full 
detector and trigger simulation. The Monte Carlo simulation samples were 
processed in the same way as for the data.  The efficiencies 
$\epsilon_{B^{+}_c}$ and $\epsilon_{B^{+}}$ 
for $p_T(B)>$ 4 (6) GeV/$c$, along with the relative efficiency, are 
presented in Table~\ref{tab:relative_eff}.
\begin{table}[htbp]
\begin{center}
\caption{Efficiencies for $B^{+}_c$ and $B^{+}$ for $p_T(B)>$ 4 (6) GeV/$c$.}
\label{tab:relative_eff}
\begin{tabular}{|l|c|c|}
\hline
Efficiency                & $p_T(B)>$ 4 GeV/$c$        & $p_T(B)>$ 6 GeV/$c$   \\
\hline					       					     
$\epsilon_{B^{+}_c}$ (\%) & $0.0551\pm 0.0010$      & $0.1232\pm 0.0024$  \\
\hline
$\epsilon_{B^{+}}$ (\%)   & $0.3231\pm 0.0022$      & $0.6005\pm 0.0042$  \\
\hline
$\epsilon_{rel}$          & $5.867\pm 0.068$ (stat) & $4.873\pm 0.060$ (stat)\\
\hline
\end{tabular}
\end{center}
\end{table}

\begin{figure}[ht]
\centering
\includegraphics[width=40mm]{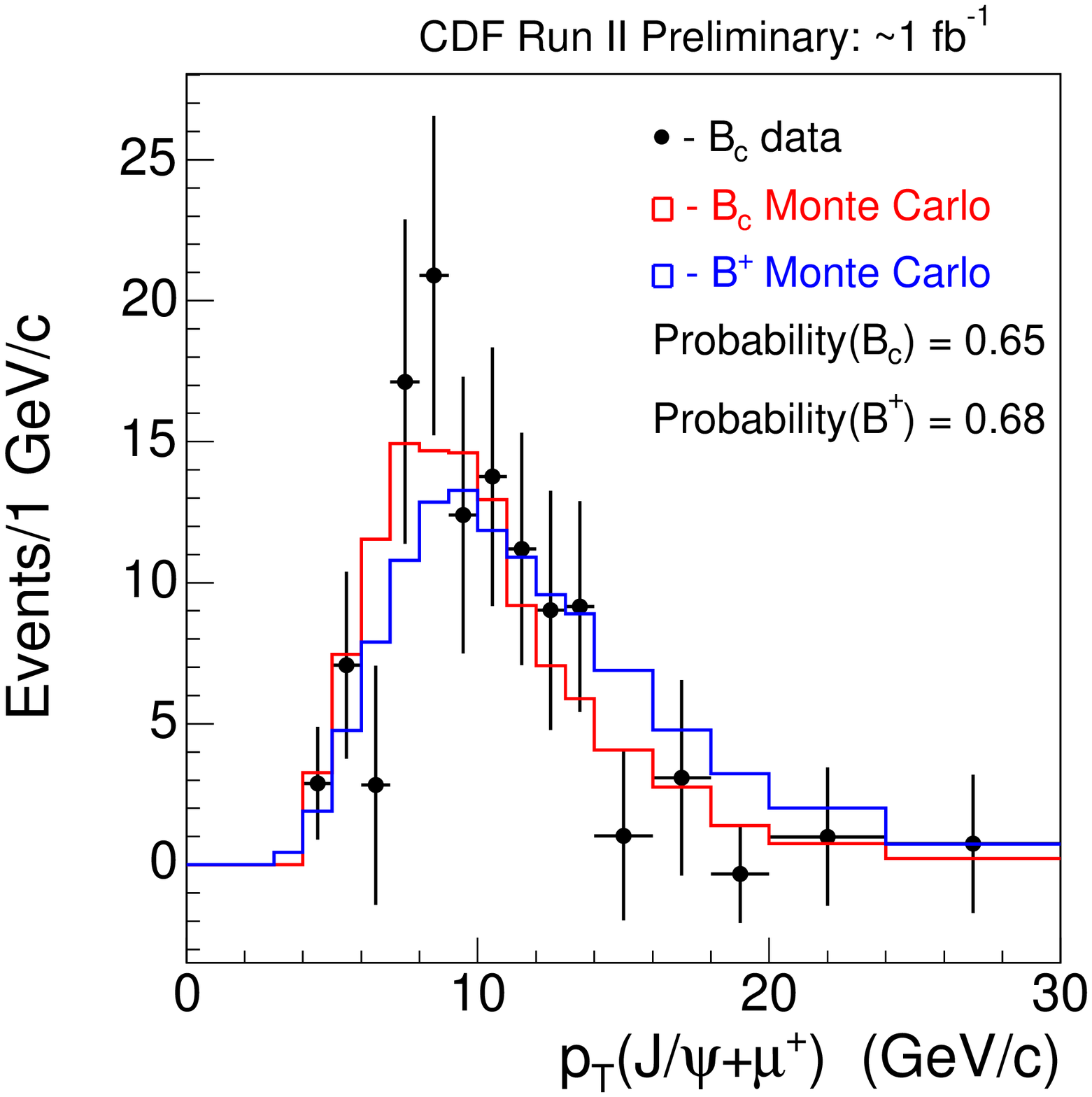}
\includegraphics[width=40mm]{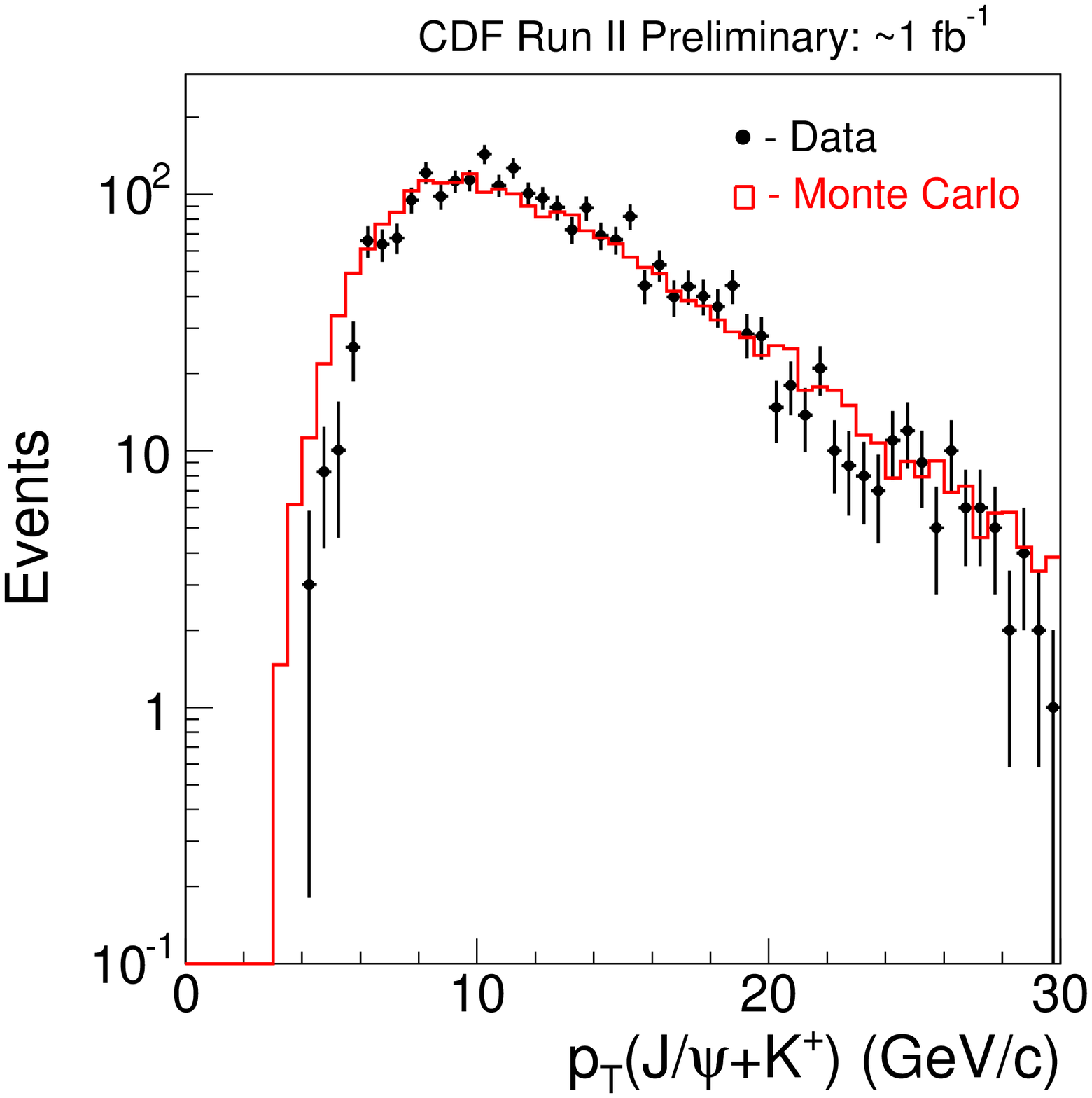} \\
\caption{The comparison of the $p_T$ spectra of data versus Monte Carlo 
simulation for $B^{+}_{c}\to J/\psi\mu^{+}\nu$ (left), and for $B^{+}\to 
J/\psi K^{+}$ decays (right).} 
\label{fig:bc_bp_ptSpec}
\end{figure}

The $p_T$ spectra for data and Monte Carlo simulation are shown in 
Fig.~\ref{fig:bc_bp_ptSpec}.

%%%%%%%%%%%%%%%%%%%%%%%%%%%%%%%%%%
\subsection{\label{sec:bc_Xsec_systematics} Cross section systematic 
uncertainties}

We divide the systematic uncertainties into two categories: uncertainties on 
the number of $B^{+}_c$ signal events and uncertainties in the determination 
of the relative efficiency.  

%%%%%%%%%%%%%%%%%%%%%%%%%%%%%%%%%%
\subsubsection{\label{sec:bc_systematics} \boldmath{$B^{+}_{c}$ background
uncertainty}}

The systematic uncertainty considered in the determination of
$N(B^{+}_{c})$ arise from events which do not originate from $B^{+}_{c}$ 
decays. All backgrounds are assigned a systematic uncertainty except the
misidentified $J/\psi$ background.  Since the estimation of this
background is determined directly from the sidebands of the $J/\psi$, we do 
not know of any source of systematic error that should be included.

The largest source of uncertainty in the misidentified third muon calculation 
is due to the poor knowledge of the proton fraction in the $J/\psi+track$ 
sample.  The particle identification method, dE/dx, does not allow us to
separate protons from kaons in our kinematic region. Consequently, we measure
the proton fraction in other momentum ranges using the time-of-flight (TOF)
particle identification combined with dE/dx information and then extrapolate 
the fraction to our momentum range according to measured trends in Monte Carlo 
simulation.

The other $N(B^+_c)$ systematic uncertainty arises from a poor knowledge 
of non-exclusive $B^+_c\to J/\psi\mu^+ X$ branching ratios and is estimated by 
varying the branching ratios of eleven $B^+_c$ decay modes that may 
contribute to the sample of tri-muon events.  In order to assign a systematic 
uncertainty
we double and halve the branching ratios of the non-exclusive decays
with respect to the rate of $B^+_c\to J/\psi\mu^+\nu$.  We choose the larger
variation in either direction for the systematic uncertainty.

Table~\ref{tab:Nbc_systematics} summarizes all of the
$B^{+}_c$ background systematic uncertainties assigned.
\begin{table}[htbp]
\begin{center}
\caption{Systematic uncertainties on the number of 
$B^{+}_{c}\to J/\psi\mu^{+}\nu$ events for different $p_T(B)$ thresholds. 
The total uncertainty is calculated by adding all the individual 
uncertainties in quadrature.}
\label{tab:Nbc_systematics}
\begin{tabular}{|l|c|c|}
\hline
$N(B^{+}_c)$ uncertainties  & $p_T(B)>$ 4 GeV/$c$ & $p_T(B)>$ 6 GeV/$c$     \\
\hline
Misid. third muon        & $\pm 5.7$ (sys)     & $\pm 5.5$ (sys)         \\
\hline
Doubly misid.            & $\pm 0.9$ (sys)         & $\pm 0.8$ (sys)      \\
\hline
Other $B^{+}_c$ decays   & $^{+6.0}_{-2.8}$ (sys)  & $^{+5.6}_{-2.5}$ (sys)  \\
\hline			          
Total                    & $^{+8.3}_{-6.4}$ (sys)  & $^{+7.9}_{-6.1}$ (sys)  \\
\hline
\end{tabular}
\end{center}
\end{table}

%%%%%%%%%%%%%%%%%%%%%%%%%%%%%%%%%%
\subsubsection{\label{sec:erel_systematics} Relative efficiency
systematic uncertainty}

We consider the systematic uncertainty in the prediction of the
relative efficiency due to the measured statistical uncertainty of the 
$B^+_c$ lifetime, knowledge of the production spectra for $B^+_c$ 
and $B^+$, and differences between $K$ and $\mu$ triggering rates at the 
first level of the CDF trigger system, the extremely fast trigger, XFT.  

The relative efficiency systematic uncertainty due to $B^+_c$ 
lifetime uncertainty is estimated by varying the lifetime by $\pm$14 
$\mu$m of its default value. This variation represents approximately one 
standard deviation in the average lifetime result.

The  relative efficiency systematic uncertainty due to the knowledge of 
$B^+_c$ $p_T$ spectrum is fully based on the theoretically predicted 
$p_T$ spectra from work in Ref.~\cite{chang}. We consider the variations 
between: 
\begin{itemize}
\item{doubling the $q\bar{q}$ contribution relative to the nominal approach;}
\item{the pure gluon fusion model, called the fixed flavor number ( FFN) 
model, and the more complete $gg+g\bar{b}+gc$ model, known as the 
general-mass variable-flavor-number (GMVFN) model;}
\item{the combined $B^+_c+B^{*+}_c$ spectrum and a pure $B^+_c$ spectrum.}
\end{itemize}

The systematic uncertainty due to knowledge of the $B^+$ $p_T$ spectrum 
is estimated by varying the Monte Carlo simulated spectrum below 10 GeV/$c$ 
to bring it into agreement with the data (see the right 
plot in Fig.~\ref{fig:bc_bp_ptSpec}. The difference between the nominal and 
recalculated relative efficiency is assigned as the uncertainty due to 
$B^+$ $p_T$ spectrum.

Another source of systematic uncertainty that we consider is the
different XFT efficiencies of kaons and muons that exist in the data
and are not modeled in the simulation.

The total $\epsilon_{rel}$ systematic 
uncertainty is summarized in Table~\ref{tab:erel_sys}.
\begin{table}[htbp]
\begin{center}
\caption{Systematic uncertainty assigned to $\epsilon_{rel}$ for different
$p_T(B)$ thresholds. The numbers 0.720 and 0.298 in the ``Total'' represent 
the uncertainties due to of $B^{+}_c$ spectrum.}
\label{tab:erel_sys}
\begin{tabular}{|l|c|c|}
\hline
$\epsilon_{rel}$ uncertainties 
                          & $p_T(B)>$ 4 GeV/$c$     & $p_T(B)>$ 6 GeV/$c$  \\
\hline 
$B^{+}_c$ lifetime        & $^{+0.393}_{-0.223}$    & $^{+0.354}_{-0.160}$ \\
\hline
$B^{+}_c$ spectrum        & $\pm$ 0.720             & $\pm$ 0.298          \\ 
\hline
$B^{+}$ spectrum          & $\pm$ 0.340             & $\pm$ 0.161          \\
\hline
XFT systematics           & $\pm$ 0.192             & $\pm$ 0.160          \\
\hline
Total        & $^{+0.554}_{-0.450}\pm$0.720 & $^{+0.420}_{-0.278}\pm$0.298 \\
\hline
\end{tabular} 
\end{center} 
\end{table}

%%%%%%%%%%%%%%%%%%%%%%%%%%%%%%%%%%
\section{\label{sec:bc_Xsec_results} The \boldmath{$B^{+}_{c}$} relative 
cross section results}

We have performed a measurement of the relative production cross
section of $B^+_c\to J/\psi\mu^+\nu$ in inclusive $J/\psi$ data with an 
integrated luminosity of 1 fb$^{-1}$.  
We have identified a sample of 229 (214) events with an estimated
background from all sources of 111$\pm$8 (107$\pm$8) events for
$p_T(J/\psi\mu^+)>$ 4 (6) GeV/$c$, respectively. The final numbers used in 
the cross section measurement, including systematic uncertainties, are
given in Table~\ref{tab:cross_section_fn}.

\begin{table}[htbp]
\begin{center}
\caption{Final numbers used in the calculation of the relative 
$B^+_c\to J/\psi\mu^+\nu$ production cross section times the branching ratio 
for two different $p_T(B)$ thresholds.}
\label{tab:cross_section_fn}
\begin{tabular}{|l|c|}
\hline
Final values        & $p_{T}(B)>$ 4 GeV/$c$                           \\
\hline\hline 
$N(B^+_c$)          & 117.6$\pm$17.2 (stat) $^{+8.3}_{-6.4}$(sys)     \\
\hline
$N(B^+$)            & 2333$\pm$ 55 (stat)                             \\
\hline
$\epsilon_{rel}$    & 5.867$\pm$0.068 (stat) $^{+0.554}_{-0.450}$(sys)\\
                    & $\pm$0.720 ($B^+_c$ spectrum) \\
\hline\hline 
Final values        & $p_{T}(B)>$ 6 GeV/$c$                            \\
\hline		   
$N(B^+_c$)          & 107.2$\pm$16.7 (stat) $^{+7.9}_{-6.1}$ (sys)     \\
\hline
$N(B^+$)            & 2299$\pm$53 (stat)                               \\
\hline
$\epsilon_{rel}$    & 4.872$\pm$0.060 (stat) $^{+0.420}_{-0.278}$ (sys)\\
                    & $\pm$0.298 ($B^+_c$ spectrum) \\
\hline
\end{tabular} 
\end{center} 
\end{table}

We give the result for the ratio
$\frac{\sigma(B^{+}_c)BF(B^{+}_c\to J/\psi\mu^{+}\nu)}{\sigma(B^{+})BF(B^{+}\to J/\psi K^{+})}$
with $p_T(B)>$ 4 GeV/$c$ thresholds as  \\

 $0.295\pm 0.040~\mbox{(stat.)}^{+0.033}_{-0.026}~\mbox{(syst.)}\pm0.036~(p_{T}~\mbox{spec})$  \\

and for $p_T(B)>$ 6 GeV/$c$ as \\

$0.227\pm 0.033~\mbox{(stat.)}^{+0.024}_{-0.017}~\mbox{(syst.)}\pm0.014~(p_T~\mbox{spec})$.
\\

Of the two results, the measurement with the $p_T(B)>$ 6 GeV/$c$
threshold has the lower systematic error.  Below 6 GeV/$c$ there is
uncertainty in the $B^+$ efficiency that appears to introduce a
significant systematic discrepancy between the simulated spectrum and
the spectrum as determined from the data.

Using theoretical assumptions and independent measurements, we are
then able to calculate the total $B^+_c$ cross section.  Using the
measured quantities $BR(B^+\to J/\psi K^+)$ = $(1.007\pm 0.035)\times
10^{-3}$~\cite{Ref:PDG} and $\sigma(B^+) = 2.78\pm 0.24~\mu$b for
$p_T(B^+)>$ 6 GeV/$c$~\cite{CDF-bplus-Xsect}, we calculate
% ++++++++++++++++++++++++++++++++++++++++++++++++++++++++++++++++++++++
\begin{eqnarray*}
  \sigma(B^+_c)*BR(B^+_c\to J/\psi\mu^+\nu) = 0.64\pm 0.20~\mbox{nb} 
\end{eqnarray*}
% ++++++++++++++++++++++++++++++++++++++++++++++++++++++++++++++++++++++
for $p_T(B^+_c)>$ 6 GeV/$c$. Assuming that the branching ratio 
$BR(B^+_c\to J/\psi\mu^+\nu) = 2.07\times
10^{-2}$~\cite{Ref:BcIvanov}, we find the total $B^+_c$ cross section 
to be
% ++++++++++++++++++++++++++++++++++++++++++++++++++++++++++++++++++++++
\begin{eqnarray*}
  \sigma(B^+_c) = 31\pm 10~\mbox{nb} 
\end{eqnarray*}
% ++++++++++++++++++++++++++++++++++++++++++++++++++++++++++++++++++++++

%%%%%%%%%%%%%%%%%%%%%%%%%%%%%%%%%%
\section{\label{sec:bc_conclusion} Conclusions}

We have performed measurements of the $B^+_c$  lifetime and production 
properties based on semileptonic $B^{+}_{c}\to J/\psi+l^{+}+X$ decays using 
data from $p\bar{p}$ collisions collected with the CDF\,II detector 
corresponding to an integrated luminosity of 1 fb$^{-1}$ at 
$\sqrt{s}$=1.96 ~TeV.

%\begin{acknowledgments}
%This document is adapted from the instructions provided to the authors
%of the proceedings papers
%\end{acknowledgments}

\bigskip % extra skip inserted
% Create the reference section using BibTeX:
%\bibliography{basename of .bib file}

\end{document}